\documentclass[final]{aipproc}

\layoutstyle{8x11single}

\SetInternalRegister\hbadness{8000} 

\newcommand\doingARLO[2][]{%
  \ifx\mmref\undefined #1\else #2\fi
}

\def\be{\begin{equation}}
\def\ee{\end{equation}}
\def\bea {\begin{eqnarray}}
\def\eea {\end{eqnarray}}
\def\ra {\rightarrow}

\begin{document}

\title{Photons from Nucleus-Nucleus Collisions at Ultra-Relativistic
Energies}

\author{Jan-e Alam}{
 address={Variable Energy Cyclotron Centre, 1/AF Bidhan Nagar, Kolkata 700 064, 
India},
}

\author{T. Hatsuda}{
  address={Department of Physics, University of Tokyo, Tokyo 113 0033, 
Japan},
}

\author{Tapan K. Nayak}{
 address={Variable Energy Cyclotron Centre, 1/AF Bidhan Nagar, Kolkata 700 064, 
India},
}

\author{Pradip Roy}{
  address={Saha Institute of Nuclear Physics, 
1/AF Bidhan Nagar, Kolkata 700 064,
India},
}

\author{Sourav Sarkar}{
 address={Variable Energy Cyclotron Centre, 1/AF Bidhan Nagar, Kolkata 700 064, 
India},
}

\author{Bikash Sinha}{
  address={Variable Energy Cyclotron Centre
and Saha Institute of Nuclear Physics, 1/AF Bidhan Nagar, Kolkata 700 064,
India},
}

\begin{abstract}
We compare the photon emission rates from hot hadronic matter
with in-medium mass shift and Quark Gluon Plasma (QGP). It is
observed that the WA98 data can be well reproduced by
hadronic initial state with initial temperature
$\sim 200$ MeV if the  universal scaling of 
temperature dependent hadronic masses are assumed
and the evolution of temperature with time is taken
from transport model or (3+1) dimensional hydrodynamics.  
The data can also be reproduced by QGP initial state
with similar  initial temperature and non-zero initial radial
velocity. 
\end{abstract}

\maketitle
\vspace*{-1cm}
\section{Introduction}
Ultra-Relativistic collisions of heavy nuclei have brought us
within reach of creating and studying various aspects 
of quark-gluon plasma (QGP), which
so far was believed to exist 
in the microsecond old universe or possibly 
in the cores of neutron or quark stars.
We are at a very interesting
situation in this area of research where the Super Proton
Synchrotron (SPS) era has drawn to a close and the first results from
the Relativistic Heavy Ion Collider (RHIC) have started to appear. 
Already from the results of the Pb run at the SPS quite a few of the 
signatures of QGP, {\it e.g.}, $J/\Psi$ suppression, strangeness enhancement
etc., are reported to have ``seen'' unmistakable hints 
of the existence of QGP~\cite{qm99}. 
Electromagnetic probes, {\it viz.}, photons
and dileptons have long been recognized as the most direct probes
of the collision~\cite{emprobe}. 
Owing to the nature of their interaction they undergo minimal scatterings
and are by far the best markers of the entire space-time evolution
of the collision.  

The single photon data obtained from Pb-Pb collisions at CERN SPS
reported by the WA98 Collaboration~\cite{wa98}
have been the focus of considerable interest in recent 
times. 
Here we emphasize the effects of in-medium modifications of hadrons
on the photon spectra 
considering the fact that as yet it has not been possible
to explain the observed low-mass enhancement of dileptons measured
in the Pb+Au as well as S+Au  collisions at the CERN SPS in 
a scenario which does not incorporate in-medium effects 
on the vector meson mass (see ~\cite{rapp} for a review).  

Let us first identify the
possible sources of ``excess'' photons above those coming
from the decays of pseudoscalar $\pi^0$ and $\eta$ mesons, as provided
by the data. Firstly, one has the prompt photons coming
from the hard collisions of initial state partons in the colliding
nuclei. These populate the high transverse momentum region 
and can be estimated by perturbative QCD.
The thermal contribution depends on the space-time evolution scenario
that one considers.
In the event of a deconfinement phase transition, one first has
a thermalized QGP which expands and cools, reverts back to hadronic matter,
again expands and cools and eventually freezes out into hadrons most of which 
are pions. Photon emission in the QGP occurs mainly due to QCD
annihilation and Compton processes between quarks and gluons.
In order to estimate the emission from the hadronic matter we
will consider a gas of light mesons {\it viz.} $\pi$, $\rho$, $\omega$,
$\eta$ and $a_1$. 

It has been emphasized by several authors that the properties  of vector mesons 
may change appreciably because
of interactions among the hadrons at high temperatures and/or densities
(see ~\cite{rev} for review).
This modifies the rate of photon emission as
well as the equation of state (EOS) of the evolving matter. 
 Among various models for vector mesons available 
 in the literature~\cite{annals}, we examine the following possibilities 
for the hadronic phase
  in this work: (i) no medium modifications of hadrons, and 
   (ii) the scenario of the universal scaling 
   hypothesis of the vector meson masses~\cite{br}.
  In principle, we can think of a third scenario (iii)
   the large collisional broadening of the vector
    mesons~\cite{rapp}.  Both (ii) and (iii) can reproduce the 
      enhancement of the low-mass dileptons measured by
       CERES Collaboration at CERN SPS, but the scenario (iii) has
been found to have a negligible effect 
on the emission rate of photons~\cite{annals}.
The effect of temperature dependent 
mass as described in case (ii) has also been incorporated
in the EOS of the hadronic matter 
undergoing a (3+1) dimensional expansion.  

There is still substantial debate on the order of the
phase transition as well as the value of the
critical temperature ($T_c$).
To address this aspect we will
also consider a scenario where
the system begins to evolve from a high temperature phase
where all the hadronic masses approach zero 
(pion mass is fixed at its vacuum value).
As the system expands and cools, the hadrons acquire
masses (as in case (ii) above) till freeze out.
Incorporation of medium modified masses and the EOS in this
case also provides a reasonable explanation of the data.

\vspace*{-0.5cm}
\section{Photon Productions}
\begin{figure}[htbp]
  \resizebox{18pc}{!}{\includegraphics[height=1.0\textheight]{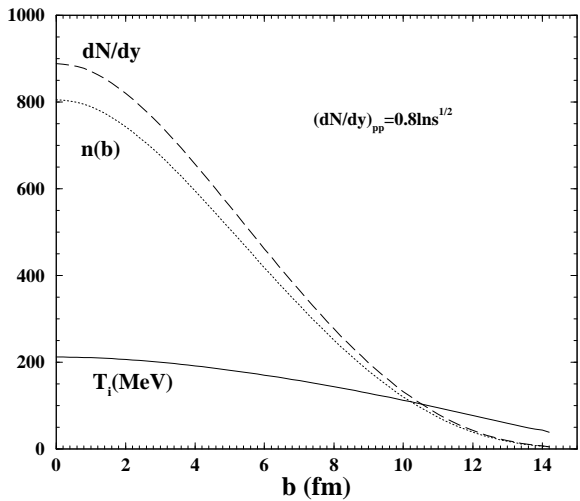}}
 \caption{
 Hadron multiplicity (long-dashed), effective number of nucleon-nucleon
collisions (dotted) and initial temperature (solid) as a function of impact
parameter calculated by using Glauber model for
nucleus - nucleus collisions at
SPS energies (see~\protect\cite{redshift} for details).} 
\label{fig1}
\end{figure}

We begin our discussions
with the prompt photon. 
The prompt photon yield for nucleus-nucleus collision is given by,
\be
E\frac{dN}{d^3p}=n(b)\frac{1}{\sigma_{in}}\,\,E\frac{d\sigma_{pp}}{d^3p}
\ee
where $n(b)$ is the average number of nucleon-nucleon 
collisions at an impact parameter $b$ ($n(b\sim 3$ fm)$\sim 660$)
as shown in Fig.~\ref{fig1}
(see~\cite{redshift} for details)  
and $\sigma_{in}\sim 30$ mb
is the $p-p$  inelastic cross section. 

The prompt photon contributions have been evaluated
with possible intrinsic transverse motion of the partons~\cite{owens,wong}
inside the nucleon and  multiplied by a $K$-factor $\sim 2$,
to account for the higher order effects. The CTEQ(5M) parton
distributions~\cite{cteq} are used for evaluating hard photons.
The relevant value of $\sqrt{s}$, energy in the centre of mass
for WA98 experiment is 17.3 GeV. No experimental data on hard
photons exist at this energy. Therefore, the ``data'' at $\sqrt{s}=17.3$
GeV is obtained from the data at $\sqrt{s}=19.4$ GeV of the E704
collaboration ~\cite{e704}
by using the scaling relation: 
 $Ed\sigma/d^3p_\gamma\mid_{h_1+h_2\,\ra\,C+\gamma}=f(x_T=2p_T/\sqrt{s})/s^2$,
for the hadronic process, $h_1+h_2\ra C+\gamma$.
This scaling is valid in the naive parton model. However,
such scaling may be spoiled in perturbative QCD due to
the reasons, among others, 
the momentum dependence of the 
strong coupling, $\alpha_s$ and from the scaling violation
of the structure functions, resulting  
in faster decrease of the cross section than
$1/s^2$. Therefore, the data at $\sqrt{s}=17.3$ GeV
obtained by using the above scaling gives a conservative
estimate of the prompt photon contributions.  The broadening
of the intrinsic transverse momentum of the partons can
play an important role both for photon~\cite{dumitru}
and neutral pion spectra~\cite{wang} in nuclear collisions.
We have neglected this effect here.

To evaluate the photon yield from quark
gluon plasma we consider the QCD Compton, annihilation,
bremsstrahlung and $q\bar q$ annihilation with scattering 
processes~\cite{kba}.
To estimate the photon yield from the hadronic matter (HM)
(see first of~\cite{kba} and \cite{npas}), 
we have considered the reactions,
$\pi\,\rho\,\ra\, \pi\,\gamma$, 
$\pi\,\pi\,\ra\, \rho\,\gamma$, $\pi\,\pi\,\ra\, \eta\,\gamma$, 
$\pi\,\eta\,\ra\, \pi\,\gamma$ and the decays 
$\rho\,\ra\,\pi\,\pi\,\gamma$
and $\omega\,\ra\,\pi\,\gamma$.
The invariant amplitudes for all these 
processes are given in Refs.~\cite{npas}.
Photon
production due to the process $\pi\,\rho\,\ra\,a_1\,\ra\,\pi\,\gamma$
is also taken into account.

 To consider the effect of the spectral modifications of hadrons
  we adopt
  two extreme cases: (i) no medium modifications of hadrons, and 
    (ii) the scaling hypothesis with  $\lambda=1/2$. 
In case (ii), the  parametrization of in-medium 
masses (denoted by $*$) at finite $T$ is

\be
{m_{V}^* \over m_{V}}  = 
 \left( 1 - {T^2 \over T_c^2} \right) ^{\lambda},
\label{anst}
\ee
where $V$ stands for vector mesons. 
Mass of the nucleon also varies with temperature
as Eq.~(\ref{anst}). In this case the width remains
constant to its vacuum value.

In (iii) the variation of the width of the vector meson ($\rho$)
with temperature is taken as, 
\be
\Gamma_{\rho}^\ast=\Gamma_\rho/(1-T^2/T_c^2), 
\label{ewidth}
\ee
and the mass remains constant to its vacuum value.
A fourth case (iv) could be the one in which 
both the mass and width of $\rho$ varies according
to Eqs.~\ref{anst} and \ref{ewidth}.

\vspace*{-0.5cm}
\section{Space-time Evolution}
We will assume that the produced matter reaches a state of thermodynamic
equilibrium after a proper time $\sim$ 1 fm/c~\cite{bj}. If a
deconfined matter is produced, it evolves in space and time till
freeze-out undergoing a phase transition to hadronic matter in
the process.  We will discuss two different models for the
description of the space time evolution:

(I) the (3+1) dimensional hydrodynamic equations 
solved numerically  by the relativistic version of the
flux corrected transport algorithm~\cite{hvg}, assuming boost
invariance in the longitudinal direction~\cite{bj} and cylindrical
symmetry in the transverse plane.
The effects of the temperature dependent hadronic masses have been
taken into account in the EOS through the effective 
statistical degeneracy~\cite{annals}.
The initial temperature $T_i$
can be related to the multiplicity of the event $dN/dy$ by virtue of the
isentropic expansion as,
\be
\frac{dN}{dy}=\frac{45\zeta(3)}{2\pi^4}\pi\,R_A^2 4a_k\,T_i^3\tau_i
\label{dnpidy}
\ee
where $R_A$ is the initial radius
of the system, $\tau_i$ is the initial thermalization time and 
$a_k=({\pi^2}/{90})\,g_k$; $g_k$ being the effective degeneracy 
for the phase $k$ (QGP or hadronic matter). The value of
$dN/dy$ is $\sim 700$ for impact parameter $b\sim 3$ fm 
(see Fig.~\ref{fig1})
The bag model EOS is used for the QGP phase. 
$g_H(T)$, the statistical degeneracy of the hadronic phase,
composed of $\pi$, $\rho$, $\omega$, $\eta$, $a_1$ and
nucleons is a temperature dependent quantity in this case 
and plays an important role in the EOS~\cite{annals}. 
As a consequence of this the square of sound velocity, 
$c_s^{-2}\,=[(T/g_H)(dg_H/dT)+3] < 1/3$, for the hadronic phase,
indicating non-vanishing interactions among the constituents.
The hydrodynamic equations have been solved with
initial energy density, $\epsilon(\tau_i,r)$~\cite{hvg}, obtained from $T_i$ 
through the EOS. We use the following relation
for the initial velocity profile,
\be
v_r=v_0\left(\frac{r}{R_A}\right)^\delta
\label{velprofile} 
\ee
For our numerical calculations we choose $\delta=1$
and sensitivity of the results on $v_0$ will be 
shown. 

(II) The integration over the space time history has also
been performed by taking the temperature profile
from the transport model~\cite{gqli},
\be
T(t)=(T_i-T_\infty)e^{-t/\tau}+T_\infty
\label{cool}
\ee
The calculation is performed for $T_i=200$ MeV, $T_\infty=120$ MeV, 
$\tau=8$ fm/c.

\vspace*{-0.5cm}
\section{Results and Discussions}
\begin{figure}[htbp]
  \resizebox{18pc}{!}{\includegraphics[height=1.0\textheight]{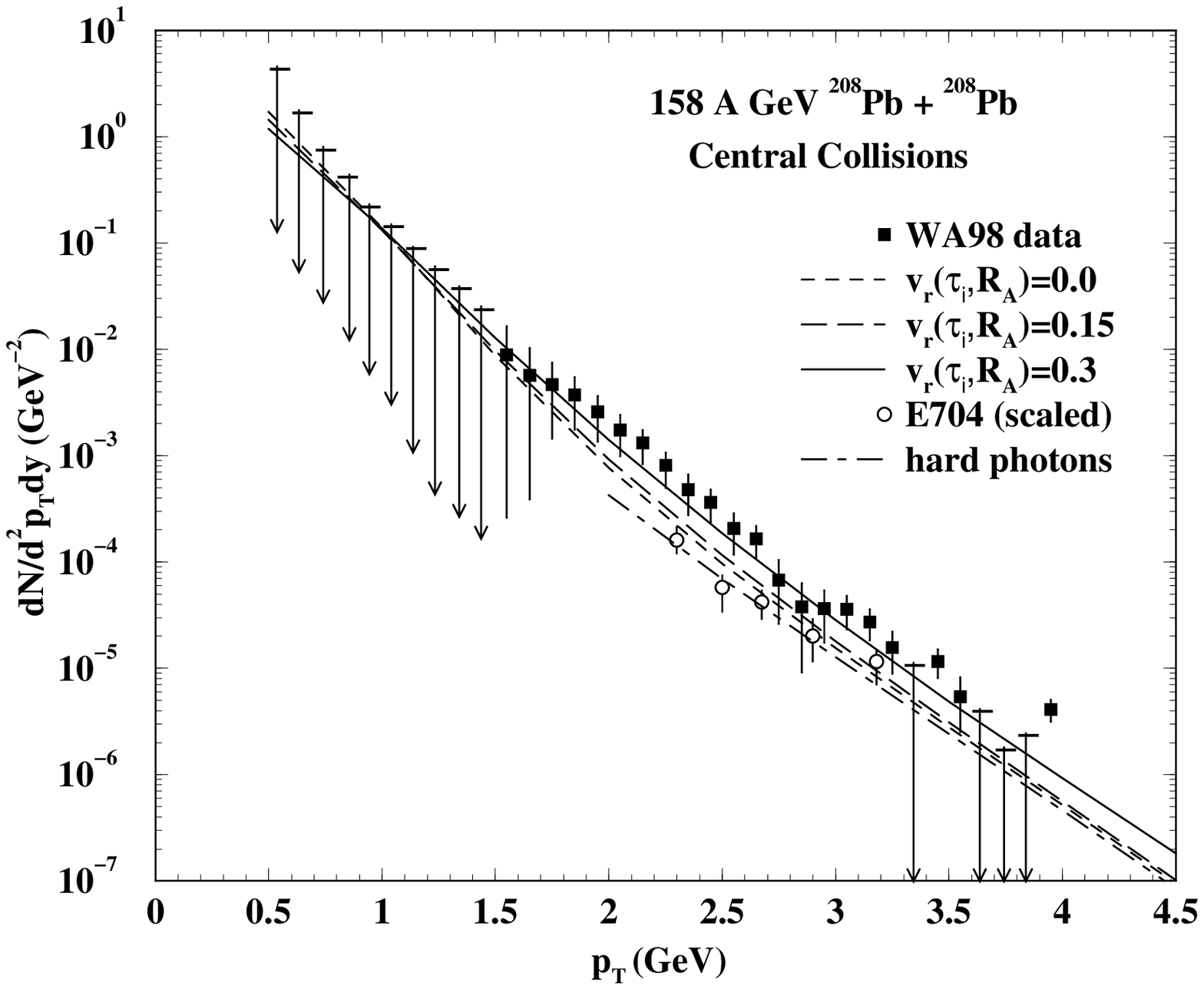}}
  \resizebox{18pc}{!}{\includegraphics[height=1.0\textheight]{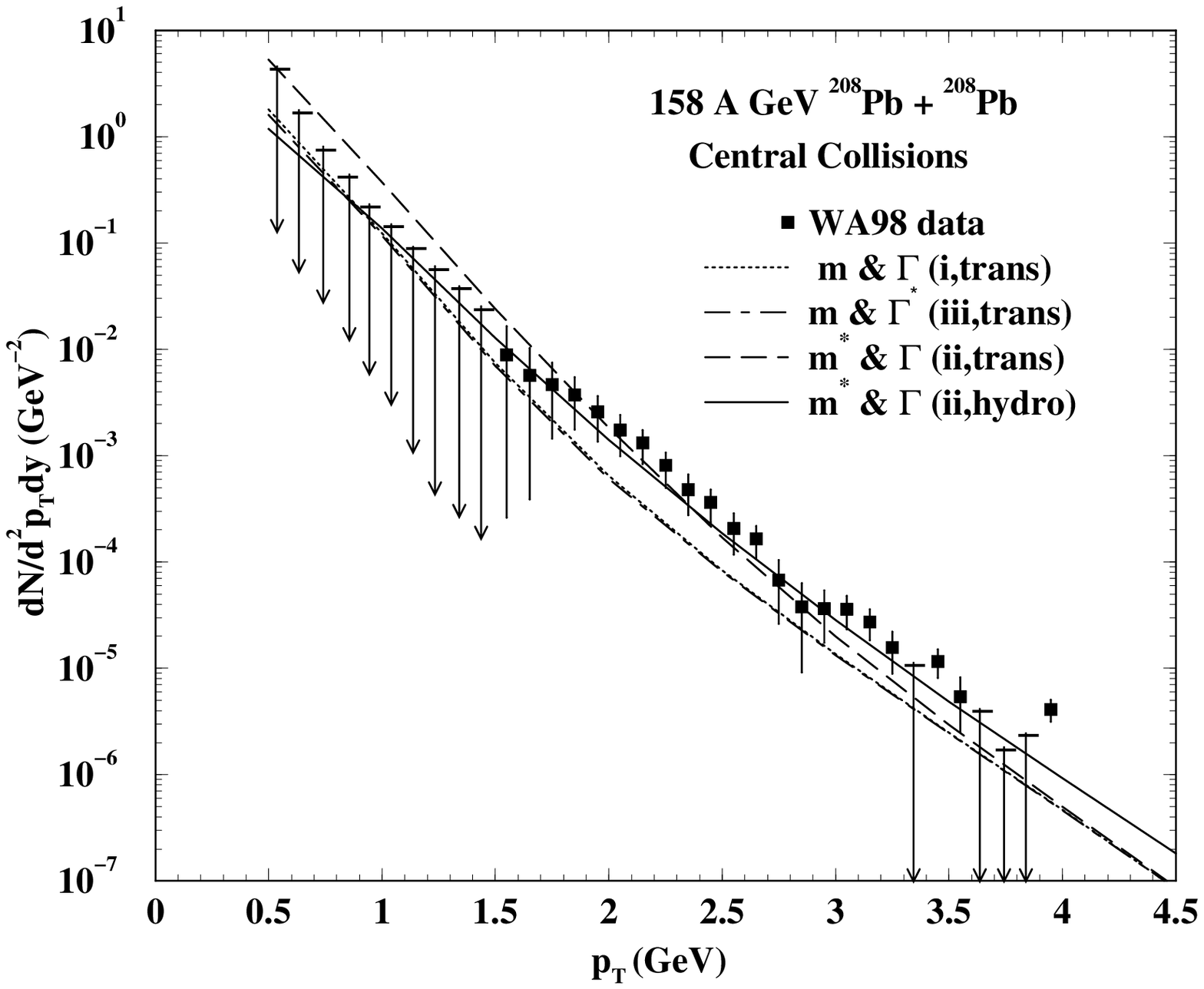}}

\caption{Left panel: Total photon yield in Pb + Pb collisions at 158 A GeV. 
The theoretical calculations contain hard QCD and thermal photons. 
The system is formed in the quark matter phase with initial temperature 
$T_i\sim 196$ MeV. Right panel: 
Total (prompt+thermal) photon yield in Pb + Pb collisions
at 158 A GeV at CERN-SPS. The theoretical
calculations contain hard QCD and thermal 
photons. 
The system is formed in the hadronic phase with initial temperature 
$T_i=200$ MeV; `trans' indicates the results 
for the cooling law~\protect(\ref{cool}).} 

\label{fig2}
\end{figure}

First we consider the QGP initial state at a temperature
$T_i\sim 196$ MeV. The values of the of the critical
temperature $T_c$ and the freeze-out temperature $T_f$
are taken as 160 and 120 MeV respectively.
In Fig.~\ref{fig2} (left),
 results for the total photon emission is shown for three different values
of the initial transverse velocity with medium effects as in case (ii).
All the three curves represent the sum of the thermal and
the prompt photon contribution which includes possible finite $k_T$
effects of the parton distributions. The later, shown separately by
the dot-dashed line also explains the scaled $pp$ data from E704 
experiment~\cite{e704}. 
We observe that the photon spectra for the initial velocity
profile given by Eq.~(\ref{velprofile}) with  $v_0=0.3$ explains
the WA98 data reasonably well. 
It is found that a substantial fraction of the photons come from mixed and 
hadronic phase. The contribution from the QGP phase
is small because of the small life time of the 
QGP ($\sim 1$ fm/c). 
    
The last statement together with the current uncertainty of the
 critical temperature $T_c$~\cite{lattice} poses the following question: Is
 the  existence of the QGP phase essential to reproduce the 
 WA98 data? To study this problem, we have considered two 
  possibilities: (a) pure hadronic model without medium-modifications,
   and (b) pure  hadronic model  with scaling hypothesis according to
    Eq.(\ref{anst}).
   In the former case,   
$T_i$ is found to be $\sim 250 $ MeV for 
$\tau_i=1$ fm/c and $dN/dy=700$,
 which appears to be too high for the hadrons to survive. Therefore
  this possibility should be excluded.
  On the other hand, 
 the second case with an assumption of $T_i = T_c$ 
(which is for simplicity) leads to 
  $T_i\sim 200$ MeV, at $\tau_i= 1$ fm/c, which is not unrealistic.
In this case, the hadronic
system expands and cools and ultimately 
freezes out at $T_f$=120 MeV. 
The masses of the vector mesons increase 
with reduction in temperature (due to expansion) according to Eq.(\ref{anst}).
The results of this scenario  
for zero initial radial velocity (including the prompt photon
contribution) are shown in Fig.~\ref{fig3} (right, solid curve). 
The experimental data is well reproduced in this case also.
This indicates that a simple hadronic model is inadequate.
 Either substantial medium modifications
   of hadrons 
  or the formation of QGP in the initial stages is necessary to
   reproduce the data. It is rather difficult to distinguish 
between the two at present.    

In Fig.~\ref{fig2} (right) the $p_T$ distribution of photons (prompt+thermal) 
is compared with the WA98 data, 
within the framework of the transport model,
the data is well reproduced 
when the hadronic masses   
are allowed to vary according to the Eqs.~\ref{anst}
(long-dash line). 
Photon spectra for scenarios
(iv) and (ii) give similar results and hence (iv) is not shown
separately.  However, when scenario (iii) is considered for thermal photons 
the experimentally observed ``excess'' photon 
in the region $1.5\le\,p_T$ (GeV) $\le 2.5$ (dashed-dot line)
is not reproduced. The dotted line indicates results
with vacuum masses and widths (scenario (i)). 

The agreement between the results obtained with two different types of 
evolution scenarios (transport model and hydrodynamics) can be explained 
as follows.
We find that the variation of temperature with time (cooling law)
in eq.~\ref{cool} is slower than the one obtained by solving 
hydrodynamic equations. As a consequence the thermal system
has a longer life time than the former case, allowing the
system to emit photons for a longer time. 
In case of hydrodynamics this is compensated by the transverse kick 
experienced by the photon at large $p_T$ due to radial velocity of the 
expanding matter.

In spite of the above encouraging situation, 
a firm conclusion about the formation of the
QGP at SPS necessitates a closer look at some pertinent
but unsettled issues. 
The hard photon contribution has been normalized
to reproduce the scaled p-p
data of E704 collaboration. However, among other
uncertainties,
the scaling we discussed before may not be valid.
It is extremely important
to know quantitatively the contribution from the hard processes.
Again, the assumption of complete thermodynamic equilibrium
for quarks and gluons may not be entirely
realistic for SPS energies; lack of chemical equilibrium
will further reduce the thermal yield from QGP. 
We have assumed $\tau_i=1$ fm/c at SPS energies, 
which may be considered as the
lower limit of this quantity,  because
the transit time (the time taken by the nuclei to pass
through each other in the CM system) is $\sim$ 1 fm/c at SPS
energies and
the thermal system is assumed to be formed after this time 
has elapsed. 
In the present work, when QGP initial state
is considered, we have assumed a first order phase transition
with bag model EOS for the QGP for its simplicity, although 
it is not in complete agreement with the lattice QCD 
simulations. 
As mentioned before, 
there are uncertainties in the value of $T_c$,
a value of $T_c\sim 200$ MeV may be considered as an
upper limit.  Moreover, the photon
emission rate from QGP considered here is obtained 
by using the hard thermal loop approximation~\cite{kba} which is 
strictly valid for $g<<1$ 
whereas the value of $g$ is $\sim 2$
at $T\sim 200 $ MeV. At present it is not clear whether the rate
obtained from HTL approximation is valid for such a large value of $g$ or not.

\begin{theacknowledgments}

J.A. is  grateful to X. N. Wang for kind hospitality at the Lawrence 
Berkeley National Laboratory where part of this manuscript was prepared.

\end{theacknowledgments}

\end{document}